\documentclass[sigconf,screen]{acmart}

\usepackage{fontawesome5}
\usepackage{soul}
\usepackage{stfloats}
\usepackage{tabularx}
\usepackage{graphicx} 
\usepackage{url} 
\usepackage{xcolor} 
\usepackage{hyperref}
\usepackage{placeins}
\usepackage{float}
\usepackage{stfloats}
 \usepackage{amsmath}

\title{A Tool for Automatically Cataloguing and Selecting Pre-Trained Models and Datasets for Software Engineering}

\author{Alexandra González, Oscar Cerezo, Xavier Franch, Silverio Martínez-Fernández}
\affiliation{%
  \institution{Universitat Politècnica de Catalunya}
  \city{Barcelona}
  \country{Spain}}
\email{{alexandra.gonzalez.alvarez, oscar.cerezo.consuegra, xavier.franch, silverio.martinez}@upc.edu}

\begin{abstract}
The rapid growth of machine learning assets has made it increasingly difficult for software engineers to identify models and datasets that match their specific needs. Browsing large registries, such as Hugging Face, is time-consuming, error-prone, and rarely tailored to Software Engineering (SE) tasks. We present \textit{MLAssetSelection}, a web application that automatically extracts SE assets and supports four key functionalities: (i) a configurable leaderboard for ranking models across multiple benchmarks and metrics; (ii) requirements-based selection of models and datasets; (iii) real-time automated updates through scheduled jobs that keep asset information current; and (iv) user-centric features including login, personalized asset lists, and configurable alert notifications. 
A demonstration video is available at \hyperlink{https://youtu.be/t6CJ6P9asV4}{https://youtu.be/t6CJ6P9asV4}.
\end{abstract}


\begin{document} 
\maketitle

\section{Introduction}
The ecosystem of Machine Learning (ML) assets \cite{zhao2024empirical}, understood as high-value artifacts contributing to ML-driven workflows such as models, datasets and metadata \cite{wang2025ml}, is vast and growing \cite{10.1145/3643991.3644898, adamenko2025swe}. As of August 2025, Hugging Face (HF) \cite{huggingface} alone surpassed two million models, establishing itself as a dominant registry for open-source ML assets \cite{laufer2025anatomy}, including top-performing and widely adopted models, with over 100k new models added each month \cite{horwitz2025we}. This growth has accelerated the reuse of ML assets, in particular in the context of Software Engineering (SE), where developers increasingly rely on existing models and datasets instead of building solutions from scratch, reducing costs, shortening development cycles, and lowering environmental impact \cite{horwitz2025we}. 
Despite these benefits, SE researchers \cite{giray2021software} and practitioners \cite{tan2024challenges, zhao2024empirical} face challenges in identifying assets meeting functional and quality requirements, as well as constraints (e.g., licensing). Choosing the right ML asset is not only a matter of performance: developers must also ensure system compatibility, avoid integration problems, and manage privacy risks, bias, and ethical concerns \cite{jiang2023empirical, tan2024challenges}. Moreover, model selection is increasingly difficult, as brute-force evaluation across model hubs is often infeasible \cite{wang2025ml}.
Consequently, despite the increasing number of SE assets \cite{gonzalez2025pretrainedmodelssupportsoftware, wang2025ml}, systematically discovering them remains challenging \cite{zhao2024empirical}, making the process time-consuming and error-prone.

To address this problem, we present \textit{MLAssetSelection}, a web tool that continuously extracts SE assets. The tool builds on our previous work on cataloguing ML models for SE \cite{gonzalez2025pretrainedmodelssupportsoftware} and the validated SEMODS dataset \cite{gonzalez2026semods}, operationalizing this classification pipeline so that SE researchers and practitioners can select and adapt \cite{zhou2025unifying} ML assets for their tasks. Its main functionalities are: (i) the \textit{Leaderboard} page, which offers a comparative view of models, where users can choose the most suitable benchmark and evaluation metric; (ii) the \textit{Models} page, where users can sample models through interactive filters; and (iii) the \textit{Datasets} page, which provides analogous filtering for datasets. The tool aims to improve the discoverability and applicability of ML assets across the entire Software Development Life Cycle (SDLC), including requirements engineering, software design, implementation, quality assurance, and maintenance. 

\textbf{Tool availability}: The tool is accessible online \footnote{\url{https://mlassetselection.essi.upc.edu}}. A replication package with the code and documentation is hosted on Zenodo \cite{gonzalez_alvarez_2025_17229471}, and a demonstration video can be found on YouTube \footnote{\url{https://youtu.be/t6CJ6P9asV4}}.

\section{Related Work}
Developing ML models from scratch is expensive and resource-intensive. The emergence of large-scale models has pushed ML engineers to adopt knowledge transfer techniques, which allow them to reuse models for downstream tasks, improving scalability and efficiency while extending applicability across diverse tasks and domains \cite{zhou2025unifying}. This practice enables organizations to spread development costs while benefiting from state-of-the-art models \cite{jiang2023empirical}. Reusing models is analogous to traditional software package reuse, but it introduces ML practices \cite{jiang2023empirical}. Recent studies highlight the challenges of reusing ML assets in SE contexts \cite{jiang2023empirical, tan2024challenges, zhao2024empirical}. Key risks include managing the model's lifecycle (e.g., missing attributes, data processing issues) and mitigating ethical and privacy concerns inherent in model deployment. 
Systematic reviews of Large Language Models (LLMs) for SE \cite{hou2024large} and classifications of open-source models for SE \cite{gonzalez2025pretrainedmodelssupportsoftware} highlight the abundance of resources while revealing a lack of systematic support for their effective selection.

Several platforms aim to centralize ML assets. HF has established itself as the dominant registry, acting as a ``model lake'' \cite{osti_10614597}, for open-source models and datasets. HF provides filtering by metadata such as licenses, libraries, languages, and ML task tags. Users can sort assets by popularity, recency, or trending status, and HF also hosts multiple leaderboards for model evaluation \cite{open-llm-leaderboard-v2, llm-perf-leaderboard}. While these features facilitate general ML asset exploration, they are often fragmented, focused on single benchmarks, and not explicitly linked to SE. For instance, HF does not provide mechanisms to map models or datasets to SE tasks such as bug prediction or code summarization. It also lacks unified exploration across multiple leaderboards and does not support reproducible exports of selected assets. 

Complementary to HF, community-driven initiatives such as HFCommunity \cite{somresearchHFCommunity, ait2023hfcommunity}  offer periodic relational database dumps of HF assets, although updates have not been released since October 2024. In parallel, Linked Papers with Code \cite{farber2023linked, linkedpaperswithcodeMetaphactory} organizes ML research as an RDF knowledge graph, linking tasks, datasets, methods, and evaluations extracted from the Papers With Code platform \footnote{In July 2025, Papers With Code was integrated into HF as \textit{Trending Papers} (https://huggingface.co/papers/trending). Although the original website is no longer available, the historical Papers With Code data remains accessible via HF datasets (e.g., https://huggingface.co/datasets/pwc-archive/papers-with-abstracts).}. However, these initiatives are designed for ML purposes and do not provide SE organization. As a result, the rapid growth of ML assets still makes it difficult for SE practitioners and researchers to identify, compare, and select those most suitable for their tasks. Prior work also shows that models hosted in dedicated ML registries are generally easier to adopt than those found in GitHub projects \cite{jiang2023empirical}, underscoring the importance of specialized tools that improve discoverability and support efficient selection of ML assets for SE.

Beyond ML registries, repository mining tools support asset discovery in SE. The SEART Data Hub \cite{dabic2024seart} is a web application that enables researchers to build and pre-process datasets of code mined from public GitHub repositories. GitHub Search \cite{dabic2021sampling} offers a dataset with 25 characteristics (e.g., commits, license) for over 735,000 GitHub repositories, alongside a web interface that allows combining selection criteria and downloading metadata of matching projects. While these tools facilitate large-scale repository exploration, they are not designed for the specific needs of SE researchers and practitioners aiming to reuse ML assets. 

In addition to registries and repository-mining tools, leaderboards have emerged to organize models by benchmarking them on standardized datasets. Examples include \textit{EvalPlus}, which compares LLM-generated code \cite{liu2023your}; the \textit{OSQ Leaderboard}, which ranks LLMs on open-style questions \cite{myrzakhan2024open, huggingfaceLeaderboardHugging}; \textit{SWE-MERA}, which evaluates LLMs for SE tasks through agentic evaluation \cite{adamenko2025swe}; and \textit{SWE-bench}, which evaluates models based on SE problems from GitHub issues \cite{jimenez2023swe}. HF also hosts various leaderboards, including some dedicated to code evaluation \cite{huggingfaceFindingBest}. While useful for assessing performance, these initiatives are fragmented, often limited to single registries, lack continuous ingestion and updates, typically focus on a single requirement (performance) \cite{ethayarajh2020utility}, and do not explain whether a model is relevant to specific SE tasks. As a result, SE practitioners lack a unified, updatable view to explore ML assets across the SDLC.

Table~\ref{tab:hf-vs-mlassetselection} summarizes the capabilities offered by HF versus \textit{MLAssetSelection}. This comparison highlights HF’s strengths (e.g., ML focus, community benchmarks) and limitations (e.g., lack of SE-task mapping, fragmented evaluations, limited filtering, minimal sorting, and absence of reproducibility). \textit{MLAssetSelection} complements HF by providing explicit SE task filters, a unified leaderboard, exportable results, and more granular filtering and sorting options relevant to SE practitioners and researchers.

\begin{table}[!h]
    \centering
    \caption{Comparison of HF and \textit{MLAssetSelection}}
    \label{tab:hf-vs-mlassetselection}
    \footnotesize
    \begin{tabular}{>{\raggedright\arraybackslash}p{2.2cm} p{2.4cm} p{2.8cm}}
    \toprule
    \textbf{Capability} & \textbf{Hugging Face} & \textbf{MLAssetSelection} \\
    \midrule
    Asset discovery scope & ML-specific & SE-focused \\ \midrule
    SE task filtering & Not supported & Explicit SE task filters \\ \midrule
    Metadata filtering & Licenses, libraries, languages, ML tasks & Licenses, libraries, languages, ML tasks, creation date \\ \midrule
    Popularity (likes, downloads) & Sort by most only; no filtering & Filterable by range; sortable ascending or descending \\ \midrule
    Activity (commits, contributors) & Not supported & Filterable by range; sortable ascending or descending \\ \midrule
    Model-specific filters & Inference providers, number of parameters, apps & Size, region, training datasets, inference providers, evaluation results \\ \midrule
    Dataset-specific filters & Size, format, modalities & Size, format, modalities, disciplines \\ \midrule
    Sorting & Trending, recency, popularity (most likes or downloads) & Name, recency, popularity (most/least likes or downloads), activity (most/least commits or contributors) \\ \midrule
    Reproducibility & Not supported & Exportable results in multiple formats (csv, json, xml) \\ \midrule
    Benchmarking & Multiple leaderboards &
    Unified leaderboard for SE \\
    \bottomrule
    \end{tabular}
\end{table}

\section{Tool Architecture}
\begin{figure}[!htb]
    \centering
    \includegraphics[width=1\linewidth]{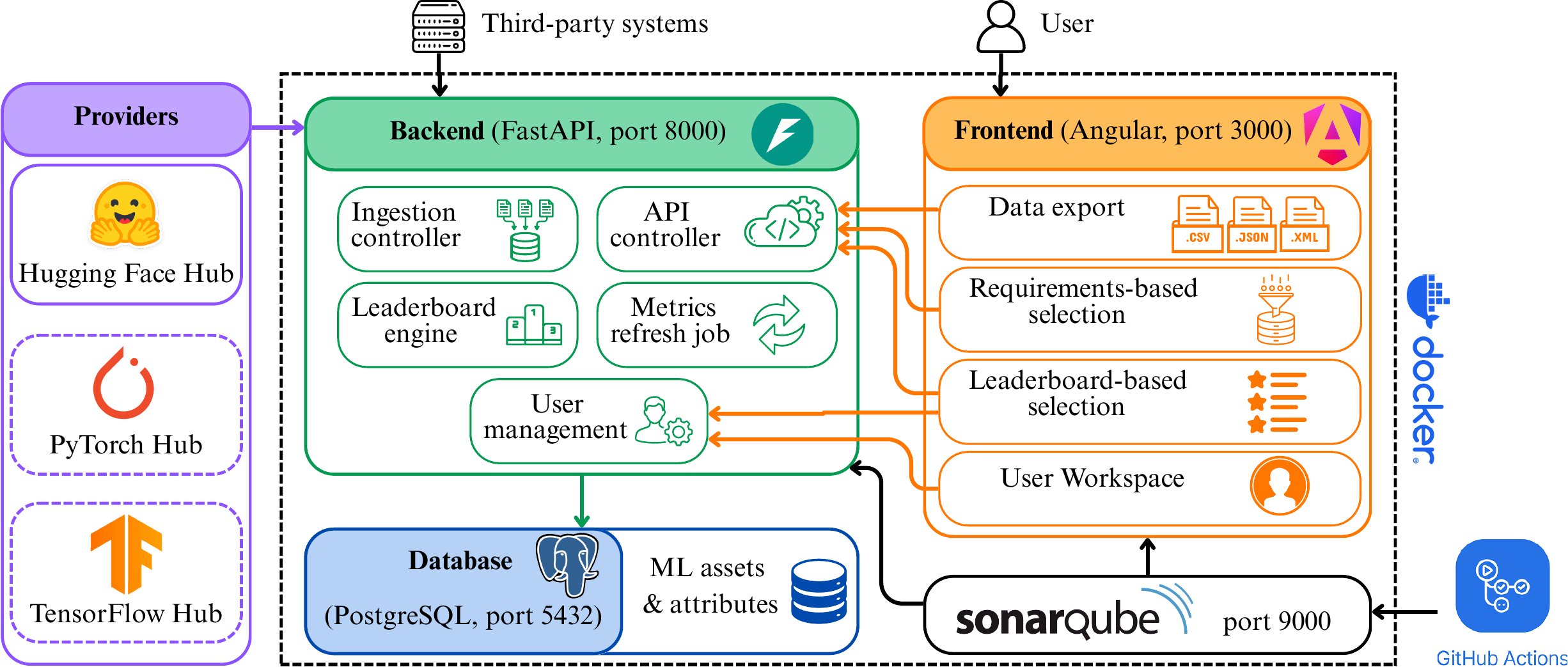}
    \caption{\textit{MLAssetSelection} architecture}
    \label{fig:arch_diagram}
\end{figure}

\textit{MLAssetSelection} follows a three-layered architecture comprising: (i) a server-side rendered Angular frontend, (ii) a FastAPI backend, and (iii) a PostgreSQL database (see Figure~\ref{fig:arch_diagram}). Designed to be provider-agnostic, the architecture facilitates the integration of ML assets from various sources. While the current implementation focuses on the HF ecosystem, the system is structured to easily plug in additional providers (e.g., PyTorch Hub, TensorFlow Hub) in future iterations. Each service runs in its own container, communicating over an isolated Docker network, while persistent volumes are mounted for database durability and static front-end assets. This setup promotes portability, enforces a clear separation of concerns, and enables one-command deployment, while allowing each tier to be scaled or replaced independently. To ensure stable code maintenance, a continuous integration and quality-gate workflow has been implemented using GitHub Actions and SonarQube. 

\subsection{Frontend}
The tool offers an interface that simplifies exploration of ML assets. Users can consult a leaderboard that ranks models across multiple benchmarks and evaluation metrics, or discover models and datasets using interactive multi-criteria requirements. In addition, filtered results can be exported in multiple file formats.

\subsubsection{Leaderboard-based Selection}
\textit{MLAssetSelection} provides a \textit{Leaderboard} ranking models across multiple benchmarks. Unlike task-specific leaderboards (e.g., those limited to code generation \cite{liu2023your, zhuo2024bigcodebench}), our approach integrates a diverse set of benchmarks, enabling users to choose the most relevant evaluation setting. Five interactive filters allow customization: (i) \textit{Benchmark}, which specifies the evaluation data (e.g., HumanEval); (ii) \textit{Implementation}, which indicates the framework or approach used (e.g., Explain); (iii) the programming \textit{Language} (e.g., C++); (iv) \textit{Metric}, which defines the performance measure (e.g., Pass@1); and (v) \textit{Metric configuration} (e.g., threshold 0.2). The leaderboard relies on self-reported performance information available in model cards: only models that report evaluation results are included, meaning not all models listed on the \textit{Models} page appear here. Based on the selected filters, users can explore performance trends (over time or relative to model size) and access the ranking. A search by model name is also available for precise queries, without affecting the ranking order.

\subsubsection{Requirements-based Selection} 
Identifying assets that meet certain requirements, such as those created within a specific time frame, is not possible directly via most registries, which typically allow basic sorting options (e.g., chronological order). Alternatives, such as community-maintained dumps \cite{ait2023hfcommunity}, may require SQL queries to join tables and determine metadata like the date of the first commit, with no straightforward way to access this information. 

With \textit{MLAssetSelection}, users can refine their search by asset \textit{identifier}, \textit{SE tasks}, \textit{metadata} (e.g., licenses, libraries, natural languages, creation date, ML tasks), \textit{popularity} (e.g., downloads, likes), \textit{activity} (e.g., commits, contributors), and attributes that are either \textit{model-specific} (e.g., size in bytes, region, training dataset, inference providers, evaluation) or \textit{dataset-specific} (e.g., size in rows, disciplines, modalities, formats). This enables users to explore repositories not only by their technical characteristics but also by their relevance to SE, without manually navigating raw data. 

By default, all assets are displayed, and users can refine the list by applying or clearing filters from a continuously updated database. Each entry links to its original repository, shows its attributes, and provides a rationale for its relevance to SE activities. Users can sort results by name, creation date, popularity, or activity metrics.

\subsubsection{Data export}
To support reproducibility and downstream analysis, \textit{MLAssetSelection} allows users to export the filtered results in standard formats (CSV, JSON, and XML). Exports include all asset attributes, so practitioners and researchers can reuse the data in workflows such as benchmarking or statistical analyses.

\subsubsection{User Workspace}
Registered users can log into the tool to access a personalized workspace. Authenticated users can save assets into customizable lists, enabling organization according to specific project needs. Additionally, users can define preferences, such as benchmarks they wish to track, and receive notifications/alerts when newly ingested models match the selected criteria.

\subsection{Backend}
The \textit{MLAssetSelection} backend runs on a web server and is responsible for managing the requests from the user interface, as well as continuously extracting newly uploaded assets to the providers and cataloguing them, retaining only those relevant to SE.

\subsubsection{Ingestion Controller}
\textit{MLAssetSelection} ingests assets from HF Hub by applying the cataloguing pipeline proposed in \cite{gonzalez2025pretrainedmodelssupportsoftware} to both models and datasets. It maps documentation (cards \cite{mitchell2019model}, metadata \cite{huggingfaceModelCardsMetadata}, and the abstract of linked arXiv papers \cite{huggingfaceModelCardsPaper}) to a taxonomy of 147 SE tasks derived from literature \cite{10.1145/3695988} and SWEBOK \cite{sommerville2015}, covering all SDLC stages. To ensure precision, the pipeline uses outlier detection to resolve lexical ambiguities and cosine-similarity to deduplicate assets. The process was validated by human experts and a large language model, achieving almost perfect agreement on Cohen's kappa ($k>0.8$) \cite{sim2005kappa}.

\subsubsection{API Controller}
The API controller serves as the interaction between the frontend and the functionalities offered by the backend. It is responsible for processing requests related to searching, filtering, and retrieving ML assets. Additionally, it enables access for third-party tools to retrieve data. The complete list of endpoints is accessible in the \textit{MLAssetSelection} documentation \cite{upcMLAssetSelectionSwagger}.

\subsubsection{Metrics Refresh Job}
Given the rapidly evolving nature of the field \cite{10.1145/3643991.3644898, myrzakhan2024open}, with a growing number of models, datasets, and benchmarks, we implemented scheduled ingestion and refresh jobs to provides state-of-the-art resources. A daily ingestion process invokes the cataloguing pipeline, following the workflow specified by González et al. \cite{gonzalez2025pretrainedmodelssupportsoftware}, to fetch newly uploaded assets. Additionally, a twelve-hourly metrics-refresh job queries the HF API to update dynamic fields (e.g., downloads, likes) of already integrated assets.

\subsubsection{Leaderboard Engine}
The leaderboard engine is responsible for extracting and processing performance metrics reported on model cards to handle inconsistencies. For all SE models detected in providers, our pipeline automatically inspects the model card metadata. We preprocess each entry to separate the reported evaluation into benchmark name, implementation, and programming language. We also extract the reported evaluation metric, along with any configuration, and its score. This process runs automatically whenever a new SE model is ingested, ensuring the leaderboard remains up-to-date and standardized without manual intervention.

\subsubsection{User Management}
The user management component handles authentication and profile management, including 2FA email verification to maximize security. It enables registered users to access personalized features such as saved assets, preference configuration, and receive alerts via the web notification centre or email. User credentials are stored securely in the database and accessed through the backend services via authenticated API requests.

\subsection{Database}
\textit{MLAssetSelection} relies on a PostgreSQL database that centralizes all models, datasets, and their associated metadata. The database is automatically populated by (i) the \textit{ingestion controller}, which integrates new models from the providers; (ii) the \textit{Leaderboard Engine}, which standardizes performance metrics to rank models; and (iii) the \textit{metrics refresh job} that continuously updates dynamic fields. Additionally, all frontend requests are handled via the \textit{API controller}, which queries the database to retrieve assets.

\section{User Workflow}
Selecting the most suitable model to augment software with AI capabilities is complex, especially for non-AI experts \cite{mueller2026assisting}. 
Consider a software engineer who wants to integrate a code generation model into a CI/CD pipeline for a C++ project. 

Without \textit{MLAssetSelection} the engineer would start by searching for models tagged with \textit{code}, resulting in thousands of candidates. Each model card would need to be inspected to determine whether it meets the requirements. To assess performance, the engineer would then consult multiple leaderboards, each focused on a single benchmark, and compare reported results that often use inconsistent metrics. As a result, the process is fragmented and time-consuming.

Using \textit{MLAssetSelection}, the workflow starts from the \textit{Leaderboard} view, where the engineer selects the target SE task (\textit{code generation}) together with the HumanEval \textit{benchmark}, the Explain \textit{implementation} variant, the C++ \textit{programming language}, and the pass@1 \textit{metric}. Based on these criteria, the tool immediately presents a ranked list of matching models. Performance trends can be explored over time or relative to model size, enabling rapid comparison of candidate models. The engineer may then export the ranking and metadata in standard formats (CSV, JSON, or XML) to document the selection process or reuse the data in downstream analyses. Users can revisit the leaderboard at any time to re-evaluate options and verify whether new models outperform their current choice. 
If additional constraints need to be applied, such as minimum popularity thresholds, licensing requirements, or activity indicators, the engineer can further refine the search using the \textit{Models} view, which provides interactive, multi-criteria filtering over the catalogue. 

Dataset selection follows a similar principle. If the engineer needs to adapt the selected model, they can navigate to the \textit{Datasets} view. Filters can be applied based on attributes such as \textit{natural language} (e.g., English), \textit{modality} (e.g., Text), and \textit{size category} (e.g., 100M–1B rows), together with popularity constraints (e.g., at least 10 \textit{downloads}). The tool returns a curated set of datasets, enabling efficient identification of suitable data without manual inspection.

To support long-term and project-oriented workflows, users can define their preferences and receive alerts when specific conditions are met (e.g., when a new model becomes available for a selected benchmark). In addition, models and datasets can be saved into user-defined lists, so that they are organized by project or experiment.

\section{Planned Evaluation}
Following the Goal Question Metric template \cite{caldiera1994goal}, we plan to evaluate \textit{MLAssetSelection} with the following \textbf{goal}: \textit{analyze} \textit{MLAssetSelection} tool \textit{for the purpose of} assessment \textit{with respect to} ease of use and usefulness \textit{from the point of} software engineers and researchers 

To obtain early feedback, we conducted a focus group in April 2025 with seven participants, including software engineering researchers and data scientists, followed by a questionnaire \cite{ToniTFG}. After a live demonstration, participants evaluated the tool using a five-point Likert scale in terms of ease of use, structural clarity, overall experience, and intention to recommend it. The results indicate strong agreement on ease of use and clarity, a positive overall experience, and a high willingness to recommend the tool. Qualitative feedback highlighted the usefulness of task-specific model discovery for SE and the ease of identifying suitable ML assets. This feedback directly informed several usability improvements, including refinements to the filters, clearer labelling, and accessibility adjustments, which are incorporated in the current version.

In the short term, we envision conducting focus groups with software engineers and researchers who regularly work with ML assets in SE contexts. Participants will interact with the tool and provide feedback on its usability and the relevance of its filters to their projects’ requirements. We will complement this with a perception-based study inspired by the Technology Acceptance Model \cite{davis1989perceived}, using questionnaires to capture participants’ views on usefulness, ease of use, and intention to adopt the tool. 

In the longer term, we plan a controlled experiment comparing practitioners using \textit{MLAssetSelection} versus direct interaction with providers, focusing on tasks such as selecting the best model for an SE task (e.g., \textit{software design}) or selecting assets with specific requirements (e.g., a minimum number of \textit{contributors}, a given range of asset \textit{sizes}), while measuring efficiency and satisfaction.

\section{Discussion}
While \textit{MLAssetSelection} streamlines the selection of SE assets, its effectiveness relies on the quality of their documentation in source registries. For instance, assets with poor documentation may not be ingested. Additionally, the \textit{Leaderboard} relies on self-reported evaluations, assuming that they were rigorously conducted. 

Regarding data privacy, the tool operates exclusively on publicly available information provided by third-party registries and respects provider API terms of service and rate limits during ingestion. \textit{MLAssetSelection} does not modify models or datasets, but catalogues them, providing direct links to the original repositories. License information, when available, is extracted from asset documentation and provided to support informed reuse decisions. 

\section{Conclusions and Future Work}
We presented MLAssetSelection, a web application that supports the selection of ML assets across the SDLC. Unlike general-purpose registries, it combines SE filtering with a standardized leaderboard, enabling users to efficiently discover and compare SE assets. 
\textit{MLAssetSelection} maintains an up-to-date catalogue that reduces manual effort and supports selection workflows. Reflecting its practical utility, the tool has already gained a base of registered users, with interest from researchers and practitioners for its adoption. 

For future work, we plan to integrate assets from additional open-source providers and marketplaces (e.g., GitHub Models \cite{githubGitHubModels}, PyTorch Hub \cite{pytorchPyTorch}). Additionally, we plan to integrate recommendation mechanisms based on user-defined preferences, enabling the discovery of relevant assets as the ML ecosystem evolves.

\begin{acks}
This work was supported by Grant PID2024-156019OB-I00 funded by MICIU/AEI/10.13039/501100011033 and by ERDF, EU. Alexandra González additionally thanks the FI-STEP grant 2025 STEP-00407.
\end{acks}

\bibliographystyle{ACM-Reference-Format}
\bibliography{references}

\appendix
\section{Tool availability for peer review}
The tool is accessible online \footnote{\url{https://mlassetselection.essi.upc.edu}}. A replication package with the code and documentation is hosted on Zenodo \cite{gonzalez_alvarez_2025_17229471}, and a demonstration video can be found on YouTube \footnote{\url{https://youtu.be/t6CJ6P9asV4}}.

\end{document}